\documentclass[twocolumn]{emulateapj}

\usepackage{graphicx}
\usepackage{epsfig}

\slugcomment{ApJ, accepted}

\shorttitle{TESTING THE NO-HAIR THEOREM: III. QUASI-PERIODIC VARIABILITY}
\shortauthors{JOHANNSEN \& PSALTIS}

\begin{document}

\title{TESTING THE NO-HAIR THEOREM WITH OBSERVATIONS IN THE ELECTROMAGNETIC SPECTRUM.\\ III. QUASI-PERIODIC VARIABILITY}

\author{Tim Johannsen\altaffilmark{1} and Dimitrios Psaltis\altaffilmark{2}}
\affil{$^1$Physics Department, University of Arizona, 1118 E. 4th Street, Tucson, AZ 85721, USA; timj@physics.arizona.edu \\
$^2$Astronomy Department, University of Arizona, 933 N.\ Cherry Ave., Tucson, AZ 85721, USA; dpsaltis@email.arizona.edu}

\begin{abstract}

According to the no-hair theorem, astrophysical black holes are uniquely described by their masses and spins. An observational test of the no-hair theorem can be performed by measuring at least three different multipole moments of the spacetime of a black hole and verifying whether their values are consistent with the unique combinations of the Kerr solution. In this paper, we study quasi-periodic variability observed in the emission from black holes across the electromagnetic spectrum as a test of the no-hair theorem. We derive expressions for the Keplerian and epicyclic frequencies in a quasi-Kerr spacetime, in which the quadrupole moment is a free parameter in addition to mass and spin. We show that, for moderate spins, the Keplerian frequency is practically independent of small deviations of the quadrupole moment from the Kerr value, while the epicyclic frequencies exhibit significant variations. We apply this framework to quasi-periodic oscillations in black-hole X-ray binaries in two different scenarios. In the case that a pair of quasi-periodic oscillations can be identified as the fundamental $g$- and $c$-modes in the accretion disk, we show that the no-hair theorem can be tested in conjunction with an independent mass measurement. If pairs of oscillations are identified with non-parametric resonance of dynamical frequencies in the accretion disk, then testing the no-hair theorem also requires an independent measurement of the black-hole spin. In addition, we argue that VLBI observations of Sgr A* may test the no-hair theorem through a combination of imaging observations and the detection of quasi-periodic variability.

\end{abstract}


\section{INTRODUCTION}

In general relativity, the no-hair theorem rests on the assumptions that any spacetime singularity must be enclosed by an event horizon (the cosmic censorship conjecture, Penrose 1969) and that the exterior spacetime is free of closed timelike curves.

If these assumptions are satisfied, astrophysical black holes are uniquely characterized by their mass $M$ and spin $J$ and are described by the Kerr metric (Israel 1967, 1968; Carter 1971, 1973; Hawking 1972; Robinson 1975). Mass and spin are the first two multipole moments of a black-hole spacetime, and all higher order moments are fully specified by the relation (Hansen 1974)
\begin{equation}
M_{l}+{\rm i}S_{l}=M({\rm i}a)^{l}.
\label{kerrmult}
\end{equation}
Here, $a\equiv J/M$ is the spin parameter, and the multipole moments are written as a set of mass multipole moments $M_l$, which are nonzero for even values of $l$, and a set of current multipole moments $S_l$, which are nonzero for odd values of $l$. Since the multipole moments of a Kerr black hole have to be related by expression (\ref{kerrmult}), the no-hair theorem can be tested by measuring (at least) three such moments (Ryan 1995).

Potential tests of the no-hair theorem have been suggested using observations of gravitational waves from extreme mass-ratio inspirals (Ryan 1995, 1997a, 1997b; Barack \& Cutler 2004, 2007; Collins \& Hughes 2004; Glampedakis \& Babak 2006; Gair et al.\ 2008; Li \& Lovelace 2008; Apostolatos et al.\ 2009; Vigeland \& Hughes 2010), from electromagnetic observations of stars in close orbits around Sgr A* (Will 2008; Merritt et al. 2010), as well as of pulsar black-hole binaries (Wex \& Kopeikin 1999).

In the first part of this series of papers (Johannsen \& Psaltis 2010a, hereafter Paper~I), we investigated a framework for measuring three multipole moments of black holes with observations in the electromagnetic spectrum. We used a quasi-Kerr metric (Glampedakis \& Babak 2006), which contains an independent quadrupole moment of the form
\begin{equation}
Q=-M\left(a^2+\epsilon M^2\right),
\label{qradmoment}
\end{equation}
where a potential deviation from the Kerr metric quadrupole is parameterized in terms of $\epsilon$. This parameterization includes the Kerr metric as the special case $\epsilon=0$.

If a measurement yields $\epsilon\neq0$, then the compact object cannot be a Kerr black hole. Within the framework of general relativity, it can only be a different type of star, a naked singularity, or else an exotic configuration of matter (see Collins \& Hughes 2004; Hughes 2006). If, however, the compact object is otherwise known to possess an event horizon and a regular spacetime, then a quadrupole deviation implies that the no-hair theorem is incorrect and that general relativity itself breaks down in the strong-field regime. We refer to black holes of this kind as ``quasi-Kerr black holes'' (Paper I; see, also, ``bumpy black holes'' Collins \& Hughes 2004; Vigeland \& Hughes 2010).

As a first astrophysical application of our framework (Johannsen \& Psaltis 2010b, Paper~II), we simulated images of such objects and demonstrated their dependence on the quadrupole moments of their spacetimes. In particular, we showed that a bright and narrow ring outside of the shadow of the black hole (hereafter the ring of light; see Johannsen \& Psaltis 2010c) with a diameter of about $10M$ (see also, e.g., Beckwith \& Done 2005) has a shape that depends uniquely on the mass, spin, quadrupole moment, and inclination of the black hole. In this paper, we show how quasi-periodic variability observed across the electromagnetic spectrum in the emission from black holes can be used to test the no-hair theorem.

Quasi-periodic oscillations (QPOs) have been observed in several galactic binaries with the Rossi X-ray Timing Explorer (RXTE; see Remillard \& McClintock 2006; see van der Klis 2006 for definitions and analysis techniques) and in Active Galactic Nuclei (AGN) with XMM-Newton (Gierli\'nski et al. 2008). While potential QPO signals in AGN are often obscured by red noise (Benlloch et al. 2001; Vaughan \& Uttley 2005, 2006), galactic black holes usually reveal much cleaner signals. In the case of galactic black holes, QPOs are transient phenomena that occur during mostly nonthermal states of the black-hole accretion disk and during state transitions; they fall into two general classes: high-frequency (about 40-450 Hz) QPOs and low-frequency (about 0.1-30 Hz) QPOs (Remillard \& McClintock 2006).

The physical origin of the observed QPOs is not well understood. In a thoroughly developed hydrodynamic model, they attributed to normal modes of oscillation trapped by general-relativistic effects in the accretion disks around black holes (see Wagoner 1999 and Kato 2001 for reviews and references therein). Expressions of these modes and the corresponding oscillation frequencies have been derived for the case of modified Newtonian potentials (Kato \& Fukue 1980; Okazaki et al. 1987; Kato 1990; Nowak \& Wagoner 1991, 1992, 1993) and in full general relativity (Perez et al. 1997; Silbergleit et al. 2001; Ortega-Rodr\'iguez et al. 2002). Since these frequencies depend primarily on the mass and the spin of the black hole and only very little on the speed of sound, quasi-periodic oscillations provide a laboratory for tests of general relativity (see Psaltis 2003, 2008 for reviews).

Of special interest are gravity modes in the equatorial plane, the so-called $g$-modes (Perez et al. 1997; see, however, Li et al. 2003), which are trapped near the inner edge of the accretion disk, as well as corrugation modes or $c$-modes (e.g., Silbergleit et al. 2001), that precess slowly around the angular momentum vector of the black hole. Both of these modes lead to a modulation of black-hole X-ray spectra; $g$-modes usually cover the largest area of the disk near the temperature maximum, while $c$-modes affect the projected area of the disk (see, e.g., Wagoner 1999). Both the $g$- and the $c$-modes are related directly to the epicyclic frequencies of particles on (nearly) circular equatorial orbits (Perez et al. 1997; Silbergleit et al. 2001). The most robust and observable mode is expected to be the axisymmetric $g$-mode, which has also been seen in various hydrodynamic simulations (Reynolds \& Miller 2009; Mao et al. 2009; Chan 2009; see, also, Wagoner 2008).

In a different approach, since QPOs have been observed in pairs with frequency ratios of $\approx3/2$ in several sources (see Remillard \& McClintock 2006), they have also been modeled as nonlinear resonances among the Keplerian and epicyclic frequencies (Klu\'zniak \& Abramowicz 2001; Abramowicz et al. 2003). The frequencies predicted by both of these models are consistent with observations and have been used in each model to constrain the spin of the black-hole X-ray binary GRO 1655-40 (Abramowicz \& Klu\'zniak 2001; Wagoner et al. 2001).

Variability has also been observed in the emission of Sgr A* in the radio, millimeter, NIR, and X-ray bands with timescales ranging from minutes to hours (e.g., Baganoff et al. 2001; Aschenbach et al. 2004; Genzel et al. 2003; Ghez et al. 2004; B\'elanger et al. 2006; Meyer et al. 2006; Yusef-Zadeh et al. 2006; Marrone et al. 2006; Hornstein et al. 2007). Possible explanations of such variability include models of density inhomogeneities (``hot spots'') orbiting around Sgr A*. Doeleman et al. (2009) investigated the prospects of ${\rm (sub-)millimeter}$ very-long baseline interferometry (VLBI) to detect periodicity in the emission of Sgr A* and to measure its spin.

In this paper, we derive expressions for the Keplerian and epicyclic frequencies of circular equatorial motion in the quasi-Kerr spacetime. We discuss the properties of these frequencies and analyze their distinct dependencies on the mass, spin, and quadrupole moment of the black-hole spacetime. In particular, we demonstrate how this formalism can be applied in order to test the no-hair theorem with the quasi-periodic variability observed from galactic black holes, AGN, and Sgr A*.

In Section~2, we derive expressions for the Keplerian frequency and the radial and vertical epicyclic frequencies in quasi-Kerr spacetimes and analyze the stability of circular orbits. In Section~3, we apply our expressions to quasi-periodic oscillations in black-hole X-ray binaries. We formulate our conclusions in Section~4.

\section{DYNAMICAL FREQUENCIES IN QUASI-KERR SPACETIME}

In this section, we systematically derive expressions for the Keplerian frequency $\Omega_{\rm \phi}$ of a particle on a circular equatorial orbit around a quasi-Kerr black hole as well as for the frequencies of small oscillations $\kappa_{\rm r}$ and $\Omega_{\rm \theta}$ in the two directions perpendicular to such an orbit. For an alternative computation of the frequencies $\Omega_{\rm \phi}$ and $\kappa_{\rm r}$, see Glampedakis \& Babak (2006).

Our starting point is the Kerr metric $g_{\rm ab}^{\rm K}$, which in Boyer-Lindquist coordinates takes the form (e.g., Bardeen, Press, \& Teukolsky 1972)
\[
ds^2=-\left(1-\frac{2Mr}{\Sigma}\right)~dt^2-\left(\frac{4Mar\sin^2\theta}{\Sigma}\right)~dtd\phi
\]
\begin{equation}
+\left(\frac{\Sigma}{\Delta}\right)~dr^2+\Sigma~d\theta^2+\left(r^2+a^2+\frac{2Ma^2r\sin^2\theta}{\Sigma}\right)\sin^2\theta~d\phi^2
\label{kerr}
\end{equation}
\noindent with
\[
\Delta\equiv r^2-2Mr+a^2,
\]
\begin{equation}
\Sigma\equiv r^2+a^2\cos^2~\theta.
\label{deltasigma}
\end{equation}

The quasi-Kerr metric $g_{\rm ab}^{\rm QK}$ in Boyer-Lindquist coordinates is given by (Glampedakis \& Babak 2006)
\begin{equation}
g_{\rm ab}^{\rm QK}=g_{\rm ab}^{\rm K}+\epsilon h_{\rm ab}.
\label{qKerr}
\end{equation}
\noindent
In contravariant form, $h^{\rm ab}$ is
\[
h^{\rm tt}=(1-2M/r)^{-1}\left[\left(1-3\cos^2\theta\right)\mathcal{F}_1(r)\right],
\]
\[
h^{\rm rr}=(1-2M/r)\left[\left(1-3\cos^2\theta\right)\mathcal{F}_1(r)\right],
\]
\[
h^{\rm \theta\theta}=-\frac{1}{r^2}\left[\left(1-3\cos^2\theta\right)\mathcal{F}_2(r)\right],
\]
\[
h^{\rm \phi\phi}=-\frac{1}{r^2\sin^2\theta}\left[\left(1-3\cos^2\theta\right)\mathcal{F}_2(r)\right],
\]
\begin{equation}
h^{\rm t\phi}=0.
\end{equation}
The functions $\mathcal{F}_{1,2}(r)$ are given in Appendix A of Glampedakis \& Babak (2006). In its full form, the quasi-Kerr metric is valid only for slowly rotating black holes with values of the spin $a\lesssim0.4M$. Note, however, that the unperturbed spacetime (i.e., $\epsilon=0$) is formally correct up to the maximum value of the spin.

Since the quasi-Kerr metric is stationary and axisymmetric, particle trajectories in this metric are characterized by three integrals of motion. For a particle with 4-momentum
\begin{equation}
p^{\rm \alpha}=\mu\frac{dx^{\rm \alpha}}{d\tau},
\end{equation}
these are its rest mass $\mu$ (which we will set equal to unity from here on), energy $E=-p_{\rm t}$, and angular momentum about the $z$-axis $L_{\rm z}=p_{\rm \phi}$.

We use the conservation of energy and axial angular momentum to express the momentum components $p^{\rm t}$ and $p^{\rm \phi}$ of a particle in the form
\begin{equation}
p^{\rm t} = -\frac{ g_{\rm \phi\phi}E + g_{\rm t\phi}L_{\rm z} }{ g_{\rm tt}g_{\rm \phi\phi} - g_{\rm t\phi}^2 },
\end{equation}
\begin{equation}
p^{\rm \phi} = \frac{ g_{\rm t\phi}E + g_{\rm tt}L_{\rm z} }{ g_{\rm tt}g_{\rm \phi\phi} - g_{\rm t\phi}^2 }.
\end{equation}

We then bring the conservation equation for the particle's rest mass,
\begin{equation}
g_{\rm \mu\nu}p^{\rm \mu}p^{\rm \nu}=-1,
\end{equation}
into the form
\[
\frac{1}{2}\left( g_{\rm rr}\dot{r}^2 + g_{\rm \theta\theta}\dot{\theta}^2 \right)
\]
\begin{equation}
= \frac{1}{2}\left[ -g_{\rm tt}(p^{\rm t})^2 - 2g_{\rm t\phi}p^{\rm t}p^{\rm \phi} - g_{\rm \phi\phi}(p^{\rm \phi})^2 - 1 \right]\equiv V_{\rm eff}
\end{equation}
with $V_{\rm eff}$ playing the role of an effective potential for the particle motion in the coordinates $r$ and $\theta$.

The radial motion in the equatorial plane can be analyzed in terms of the equation
\begin{equation}
\frac{1}{2} \left( \frac{dr}{dt} \right)^2 = \frac{ V_{\rm eff} }{ g_{\rm rr}(p^{\rm t})^2 } \equiv V_{\rm eff}^{\rm r},
\label{Veffr}
\end{equation}
while the $\theta$-motion (for constant radii $r$) is governed by the equation
\begin{equation}
\frac{1}{2} \left( \frac{d\theta}{dt} \right)^2 = \frac{ V_{\rm eff} }{ g_{\rm \theta\theta}(p^{\rm t})^2 } \equiv V_{\rm eff}^{\rm \theta}.
\label{Vefftheta}
\end{equation}

In Paper I, we solved the equations
\begin{equation}
V_{\rm eff}^{\rm r} = 0
\end{equation}
and
\begin{equation}
\frac{d  V_{\rm eff}^{\rm r} }{dr} = 0,
\end{equation}
which characterize circular equatorial orbits, and obtained the energy
\[
E=\frac{ r^{3/2}-2Mr^{1/2}\pm aM^{1/2} }{ r^{3/4}\sqrt{r^{3/2}-3Mr^{1/2}\pm2aM^{1/2}} } 
\]
\[
- \frac{ 5\epsilon }{ 32M^2r^{3/2}(r-3M)^{3/2} }
\]
\[
\times \bigg[ 2M(6M^4+14M^3r-41M^2r^2+27Mr^3-6r^4)
\]
\begin{equation}
\left. +r^2(6r^3-33Mr^2+66M^2r-48M^3)\ln\left( \frac{r}{r-2M} \right) \right]
\label{energy}
\end{equation}
and axial angular momentum
\[
L_{\rm z}=\pm\frac{ M^{1/2}(r^2\mp2aM^{1/2}r^{1/2}+a^2) }{ r^{3/4}\sqrt{r^{3/2}-3Mr^{1/2}\pm2aM^{1/2}} }
\]
\[
\mp \frac{ 5\epsilon }{ 32M^{5/2}(r-3M)^{3/2} }
\]
\[
\times \bigg[ 2M(6M^4-7M^3r-16M^2r^2+12Mr^3-3r^4) 
\]
\begin{equation}
\left. +3r(r^4-5Mr^3+9M^2r^2-2M^3r-6M^4)\ln\left( \frac{r}{r-2M} \right) \right].
\label{lz}
\end{equation}
These expressions are expansions to linear order in the parameter $\epsilon$, where we have neglected terms of the order $\epsilon a$. An expansion of this type is implicitly understood for all expressions throughout the paper in accordance with the form of the metric specified by expression (\ref{qKerr}). Note, however, that we do not expand in the spin parameter $a$ so that our results are correct for arbitrary values of the spin in the special case $\epsilon=0$.

\section*{2.1~~THE KEPLERIAN FREQUENCY $\Omega_{\rm \phi}$}

For a particle moving on a circular orbit in the equatorial plane, the time and azimuthal components of the 4-momentum take the form (Glampedakis \& Babak 2006)
\[
p^{\rm t}=\frac{1}{\Delta} \left[ E(r^2+a^2)+\frac{2Ma}{r}(aE-L_{\rm z}) \right]
\]
\begin{equation}
 - \epsilon \left( 1-\frac{2M}{r} \right)^{-1}f_3(r)E,
\end{equation}
\begin{equation}
p^{\rm \phi}=\frac{1}{\Delta} \left[ \frac{2M}{r}(aE-L_{\rm z})+L_{\rm z} \right] - \epsilon\frac{h_3(r)}{r^2}L_{\rm z},
\label{uisco}
\end{equation}
where
\[
f_3(r)=-\frac{5(r-M)}{8Mr(r-2M)}(2M^2+6Mr-3r^2)
\]
\[
-\frac{15r(r-2M)}{16M^2}\ln \left( \frac{r}{r-2M} \right),
\]
\[
h_3(r)=\frac{5}{8Mr}(2M^2-3Mr-3r^2)
\]
\begin{equation}
+\frac{15}{16M^2}(r^2-2M^2)\ln \left( \frac {r}{r-2M} \right).
\end{equation}

From these expressions, we obtain the Keplerian frequency
\[
\Omega_{\rm \phi} \equiv \frac{d\phi}{dt} = \frac{ p^{\rm \phi} }{ p^{\rm t} }
\]
\[
= \pm\frac{ \sqrt{M} }{ r^{3/2} \pm a\sqrt{M} } \mp \frac{ 5\epsilon }{ 32M^{5/2}(r-2M)r^{5/2} }       
\]
\[
\times \bigg[ 2M(-6M^4 + 8M^3 r + 2M^2 r^2 + 3M r^3 - 3 r^4)
\]
\begin{equation}
 + 3r(r-2M)(r^3-2M^3)\ln\left( \frac{ r }{ r-2M } \right) \bigg].
\end{equation}

\begin{figure}[h]
\begin{center}
\includegraphics[width=0.45\textwidth]{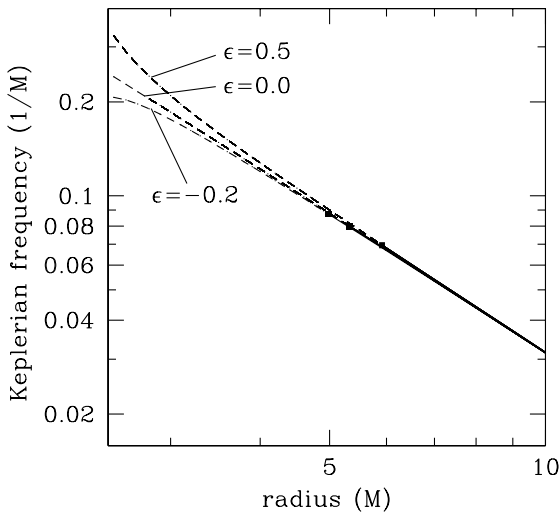}
\end{center}
\caption{Keplerian frequency $\Omega_{\rm \phi}$ of a particle on a circular equatorial orbit around a quasi-Kerr black hole with a spin of $a=0.2M$ and for several values of the parameter $\epsilon$. The Keplerian frequency increases for decreasing values of the radius and for increasing values of the parameter $\epsilon$. The dots mark the location of the ISCO, and the dashed lines correspond to unstable orbits.}
\label{omegaphifig}
\end{figure}

In Figure~\ref{omegaphifig} we plot the Keplerian frequency as a function of radius for several values of the parameter $\epsilon$ and a spin of $a=0.2M$. The frequency increases at smaller radii. The dependence on the quadrupolar parameter $\epsilon$ is weak and only manifests at radii comparable to a few $M$. Increasing the value of the parameter $\epsilon$ leads to larger Keplerian frequencies.

\section*{2.2~~THE RADIAL AND VERTICAL EPICYCLIC FREQUENCIES}

In order to calculate the radial and vertical epicyclic frequencies, we introduce small perturbations $\delta r$ and $\delta \theta$ around a circular equatorial orbit located at radius $r_0$. We take the (coordinate) time derivative of equations (\ref{Veffr}) and (\ref{Vefftheta}) and obtain
\begin{equation}
\frac{d^2(\delta r)}{dt^2} = \frac{ d^2 V_{\rm eff}^{\rm r} }{ dr^2 } \delta r,
\end{equation}
\begin{equation}
\frac{d^2(\delta \theta)}{dt^2} = \frac{ d^2 V_{\rm eff}^{\rm \theta} }{ d\theta^2 } \delta \theta
\end{equation}
using the effective potentials $V_{\rm eff}^{\rm r}$ and $V_{\rm eff}^{\rm \theta}$ defined in equations (\ref{Veffr}) and (\ref{Vefftheta}). From these expressions, we derive the frequencies of small oscillations in the $r$ and $\theta$ directions around a circular equatorial orbit as
\begin{equation}
\kappa_{\rm r}^2 = -\frac{ d^2 V_{\rm eff}^{\rm r} }{dr^2},
\end{equation}
\begin{equation}
\Omega_{\rm \theta}^2 = -\frac{ d^2 V_{\rm eff}^{\rm \theta} }{ d\theta^2 }.
\end{equation}
Here the second derivatives of the effective potentials are evaluated at the energy and axial angular momentum given by expressions (\ref{energy}) and (\ref{lz}), and for $\theta=\pi/2$. The result is:
\[
\kappa_{\rm r}^2 = \frac{ M(r^2 - 6Mr \pm 8a\sqrt{M}\sqrt{r} -3a^2) }{ r^2(r^{3/2} \pm a\sqrt{M} )^2 }
\]
\[
- \frac{ 5\epsilon }{ 16M^2 r^5(r-2M) }\times \bigg[ 2M
\]
\[
 (48M^5 + 30M^4 r +26M^3r^2 -127M^2r^3 + 75Mr^4 - 12r^5)
\]
\begin{equation}
\left. + 3r^2(r-2M)^2(4r^2 - 13Mr - 2M^2)\ln\left( \frac{ r }{ r-2M } \right) \right]
\label{kappasquared}
\end{equation}
and
\[
\Omega_{\rm \theta}^2 = \frac{ M(r^2 \mp 4a\sqrt{M}\sqrt{r} +3a^2) }{ r^2(r^{3/2} \pm a\sqrt{M})^2 }
\]
\[
+ \frac{ 5\epsilon }{ 16M^2 r^4(r-2M) }
\]
\[
\times \bigg[ 2M( 6M^4 + 34M^3r - 59M^2r^2 + 33Mr^3 - 6r^4 )
\]
\begin{equation}
\left. + 3r(2r-M)(r-2M)^3\ln\left( \frac{ r }{ r-2M } \right) \right].
\end{equation}
To convert these expressions to cgs units, one needs to multiply by $(c^3/G)^2$, where $c$ and $G$ are the speed of light and the gravitational constant, respectively. In the special case where $\epsilon=0$, these expressions coincide with the ones in Okazaki et al. (1987) and Kato (1990). Similar calculations have also been performed by Shibata \& Sasaki (1998), Berti \& Stergioulas (2004), Gair et al. (2008), and Sanabria-G\'omez et al. (2010).

\begin{figure}[h]
\begin{center}
\includegraphics[width=0.45\textwidth]{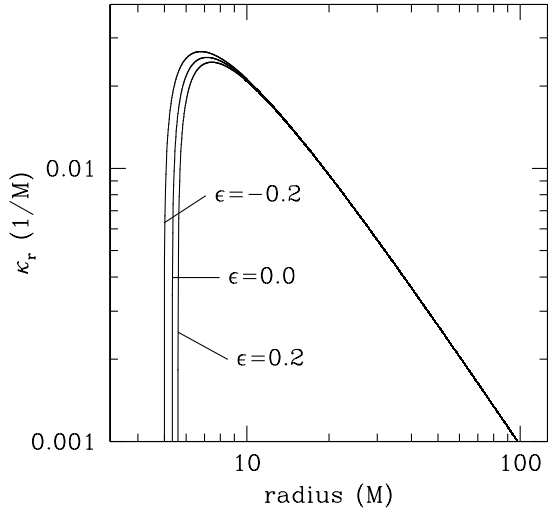}
\end{center}
\caption{Radial epicyclic frequency $\kappa_{\rm r}$ versus radius for a particle orbiting a quasi-Kerr black hole with a spin of $a=0.2M$ and for several values of the parameter $\epsilon$. The radial epicyclic frequency increases with decreasing values of the radius and reaches a maximum at $r\approx7M$ for this spin. Decreasing values of the parameter $\epsilon$ shift the maximum to smaller radii.}
\label{omegarfig}
\end{figure}

\begin{figure}[h]
\begin{center}
\includegraphics[width=0.49\textwidth]{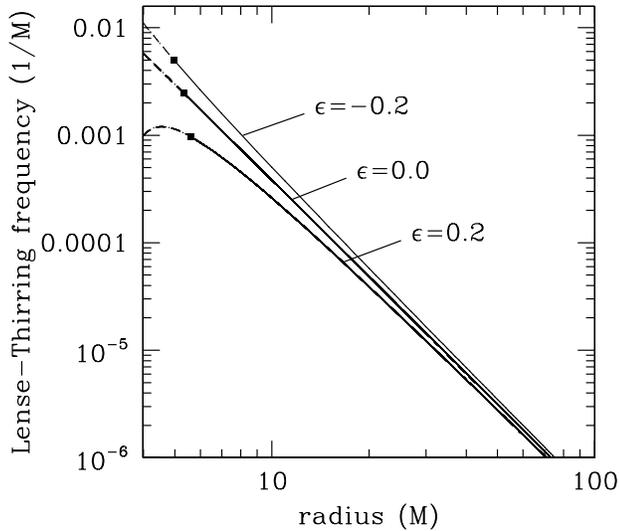}
\end{center}
\caption{Lense-Thirring frequency $\Omega_{\rm LT}=\Omega_{\rm \phi}-\Omega_{\rm \theta}$ versus radius for a particle on an orbit around a quasi-Kerr black hole with a spin of $a=0.2M$ and for several values of the parameter $\epsilon$. The Lense-Thirring frequency increases with decreasing values of the radius and of the parameter $\epsilon$. The dots mark the location of the ISCO, and dashed lines correspond to unstable orbits.}
\label{omegalthfig}
\end{figure}

We plot the radial epicyclic frequency $\kappa_{\rm r}$ as a function of radius for several values of the parameter $\epsilon$ and a spin of $a=0.2M$ in Figure~\ref{omegarfig}. The radial oscillation frequency increases with decreasing values of the radius and reaches a maximum at $r\approx7M$ for this spin. Decreasing values of the parameter $\epsilon$ shift the maximum to smaller radii.

In Figure~\ref{omegalthfig} we plot the Lense-Thirring frequency
\begin{equation}
\Omega_{\rm LT}\equiv\Omega_{\rm \phi}-\Omega_{\rm \theta}
\label{omegalth}
\end{equation}
as a function of radius for several values of the parameter $\epsilon$ at a spin value $a=0.2M$. The Lense-Thirring frequency describes the precession of the orbital plane of a particle moving around the black hole and vanishes in the case of a Schwarzschild black hole, where $\Omega_{\rm \phi}=\Omega_{\rm \theta}$. The Lense-Thirring frequency increases with decreasing values of the radius and of the parameter $\epsilon$. This frequency depends significantly on the parameter $\epsilon$ and changes by a factor of $\sim5$ at the ISCO for values of the parameter $\epsilon=\pm0.2$. Note that both the Lense-Thirring frequency and the radial oscillation frequency are significantly smaller than the Keplerian frequency $\Omega_{\rm \phi}$.

\section*{2.3~~STABLE ORBITS IN QUASI-KERR SPACETIME}

\begin{figure}
\begin{center}
\includegraphics[width=0.45\textwidth]{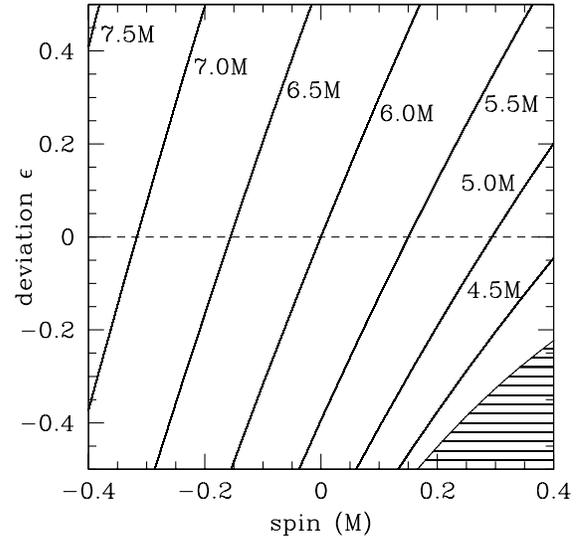}
\end{center}
\caption{Location of the ISCO as a function of the spin $a$ and the parameter $\epsilon$. The radius of the ISCO decreases with increasing values of the spin and decreasing values of the parameter $\epsilon$. The dashed line corresponds to the special case of the Kerr metric, while the shaded region is excluded and marks the part of the parameter space where higher order terms in the radial epicyclic frequency become important.}
\label{iscos}
\end{figure}

\begin{figure}
\begin{center}
\includegraphics[width=0.45\textwidth]{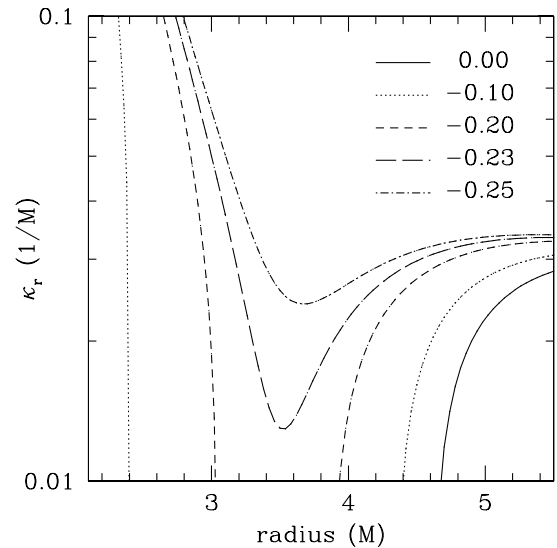}
\end{center}
\caption{Radial epicyclic frequency $\kappa_{\rm r}$ as a function of radius for a black hole of spin $a=0.4M$ and for several values of the parameter $\epsilon$. Zeros of the radial epicyclic frequency correspond to marginally stable orbits. Small negative values of the parameter $\epsilon$ lead to the emergence of a second marginally stable orbit in addition to the ISCO. For values of the parameter $\epsilon\lesssim -0.23$, these two merge and all orbits are now stable against radial perturbations. The ISCO is then determined by the onset of a vertical instability.}
\label{veff}
\end{figure}

\begin{figure}
\begin{center}
\includegraphics[width=0.45\textwidth]{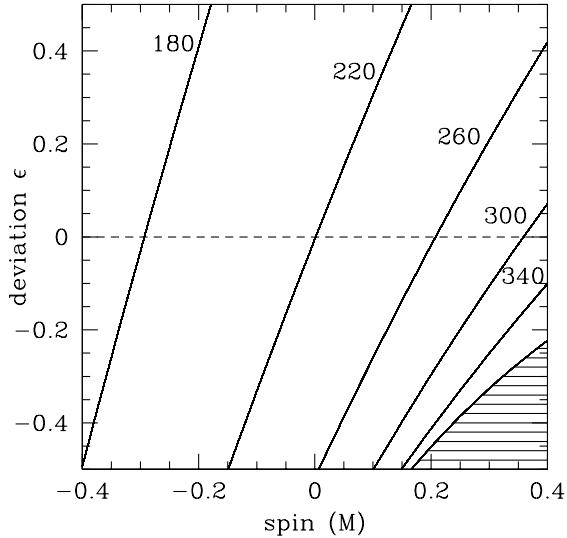}
\end{center}
\caption{Maximal Keplerian frequency in Hz scaled appropriately for a $10M_\odot$-mass black hole as a function of the spin $a$ and the parameter $\epsilon$. The maximal Keplerian frequency increases with increasing values of the spin and decreasing values of the parameter $\epsilon$. The dashed line corresponds to the special case of the Kerr metric. The shaded region marks the excluded part of the parameter space where higher order terms in the radial epicyclic frequency become important.}
\label{ophimax}
\end{figure}

In this section, we determine the location of the ISCO in the quasi-Kerr spacetime which separates the region of space where circular orbits are stable from the region where circular orbits are unstable. The location of this orbit, in turn, determines the maximum Keplerian frequency for that particular central object.

In Paper~I we found the location of the ISCO from equation (\ref{energy}) by solving
\begin{equation}
\frac{dE}{dr}=0.
\label{Eeqisco}
\end{equation}
In this paper, we determine the location of the ISCO by computing the radius $r$ at which the frequency of small radial oscillations $\kappa_{\rm r}$ becomes imaginary, rendering circular equatorial motion unstable (see, also, Shibata \& Sasaki 1998; Berti \& Stergioulas 2004; Gair et al. 2008).

In Figure~\ref{iscos} we plot the location of the ISCO as a function of the spin and the parameter $\epsilon$. The location of the ISCO shifts to smaller radii for increasing values of the spin as in the case of Kerr black holes, while the radius of the ISCO increases for increasing values of the parameter $\epsilon$. We denoted the location of the ISCO in Figures~\ref{omegaphifig} and \ref{omegalthfig} by dots.

In addition to the region outside of the ISCO, we find a second branch of stable circular orbits similar to the ones reported in Gair et al. (2008). If the parameter $\epsilon$ is sufficiently negative for a given value of the spin, we observe the emergence of this branch at radii smaller than the radius of the ISCO as well as of a second unstable branch for which the epicyclic frequency $\Omega_{\rm \theta}$ becomes imaginary.

We illustrate this behavior in Figure~\ref{veff}, where we plot the radial epicyclic frequency as a function of radius for a black hole with spin $a=0.4M$ and for several values of the parameter $\epsilon$. For a Kerr black hole, the radial epicyclic frequency has only one zero at a radius $r_{\rm ISCO}\approx4.6M$. If $\epsilon\lesssim0$, the location of the ISCO decreases, while a second branch of stable circular orbits emerges in the neighborhood of the circular photon orbit. The radius of the second marginally stable circular orbit increases with decreasing values of the parameter $\epsilon$. For values of the parameter $\epsilon\lesssim-0.23$, the radial epicyclic frequency becomes strictly positive and these two marginally stable circular orbits disappear. In this case, all circular orbits are stable against radial perturbations, and the ISCO is determined by instability in the vertical direction.

Gair et al. (2008) referred to the larger one of these additional two marginally stable circular orbits as OSCO, the outermost stable circular orbit. In all cases, however, $r_{\rm OSCO}<r_{\rm ISCO}$. Since a particle on a circular orbit in this region has a larger energy and axial angular momentum than a particle on a circular orbit in the region $r\geq r_{\rm ISCO}$, the region bound by the OSCO, if present, is likely to be vacant of particles on circular orbits (see discussion in Gair et al. 2008). Generally, however, this region is very close to the black hole, where the quasi-Kerr metric is no longer perturbative (see Paper I).

The shaded region in Figure~\ref{iscos} marks the part of the parameter space where all circular equatorial orbits are radially stable. Albeit interesting in its own merit, we do not further consider this part of the parameter space, because higher order terms in the radial epicyclic frequency become important.

For particles in an accretion disk, circular motion at radii smaller than the ISCO is unstable, and the Keplerian frequency reaches its maximal value at the ISCO. The maximum Keplerian frequency is the largest oscillation frequency in the disk and, therefore, an upper bound on the dynamical frequencies. In Figure~\ref{ophimax}, we plot the maximal Keplerian frequency $\Omega_{\rm \phi,max}$ as a function of the spin $a$ and the parameter $\epsilon$. Since the Keplerian frequency depends only weakly on the parameter $\epsilon$ (see Figure~\ref{omegaphifig}), quadrupole deviations affect the maximal Keplerian frequency predominantly due to the shift of the ISCO. The shaded region, again, marks the excluded part of the parameter space where higher order terms in the radial epicyclic frequency become important.

\section{Quasi-Periodic Oscillations}

In this section, we use our expressions of the Keplerian and epicyclic frequencies that we derived in Section~2 to demonstrate how QPOs observed in black-hole X-ray binaries can be used to test the no-hair theorem in two different scenarios. In both interpretations of detected QPO signals, the dynamical frequencies are important but play different roles. We argue that either one or two of the three parameters mass, spin, and quadrupole moment can be determined from a pair of QPOs and that the no-hair theorem can be tested in conjunction with an independent measurement of the remaining parameters.

\section*{3.1~~THE DISKOSEISMOLOGY MODEL}

In general relativity, modes can be trapped in the accretion disk of a black hole giving rise to modulations of the measured flux. Detailed expressions of these modes have been derived in a hydrodynamic model in full general relativity (the diskoseismology model; Perez et al. 1997; Silbergleit et al. 2001). In this interpretation, the observed QPO pair can be identified as the lowest order gravity ($g$-modes) and corrugation modes ($c$-modes). The frequencies of these modes depend almost exclusively on the mass and spin in the case of a Kerr black hole and, consequently, both mass and spin can be determined from a pair of QPOs (e.g., Wagoner et al. 2001).

The radial epicyclic frequency $\kappa_{\rm r}$ has a maximum near the inner edge of the accretion disk (Okazaki et al. 1987). In Kerr geometry, the lowest order $g$-modes occur at frequencies that are very close to that maximum (Perez et al. 1997). Corrugation modes, on the other hand, can exist only in disks that corotate with the Kerr black hole with a spin $a$ in the range $0<a_0\leq a\lesssim0.95$, where $a_0\sim10^{-5}-10^{-3}$ (Silbergleit et al. 2001). The fundamental $c$-mode coincides with the Lense-Thirring frequency of the black hole evaluated at a radius $r_{\rm c}$ which is typically very close to the ISCO. For a black hole with a mass of $10M_\odot$ and a disk luminosity $L\sim10\%$ of the Eddington luminosity, Silbergleit et al. (2001) estimate the radius $r_{\rm c}$ by the formula
\begin{equation}
r_{\rm c}=r_{\rm ISCO} + K_0 a^{-K_1}(1-a)^{K_2},
\label{rc}
\end{equation}
where $K_0=0.093M$, $K_1=0.79$, and $K_2=0.20$.

In a quasi-Kerr spacetime, the frequencies of the $g$- and $c$-modes depend not only on the mass and spin of the black hole, but also on the quadrupole moment through the parameter $\epsilon$. Therefore, the no-hair theorem can be tested if either the mass or spin is measured independently, as we demonstrate in the following.

For our analysis, we make the simplifying assumptions that the frequency of the fundamental $g$-mode is equal to the maximum of $\kappa_{\rm r}/2\pi$ and that the fundamental $c$-mode is equal to the Lense-Thirring frequency of a quasi-Kerr black hole evaluated at the ISCO. For a Kerr black hole, the first assumption is an excellent approximation (Perez et al. 1997), while the second assumption provides an estimate that slightly deviates from the true value $\Omega_{\rm LT}(r_{\rm c})/2\pi$. In this working definition, we also assume that quasi-Kerr $c$-modes exist as long as $\Omega_{\rm LT}(r_{\rm ISCO})>0$. For a rigorous treatment, a detailed study of perturbations on accretion disks in quasi-Kerr spacetime is necessary, which is beyond the scope of this paper.

\begin{figure}[ht]
\begin{center}
\psfig{figure=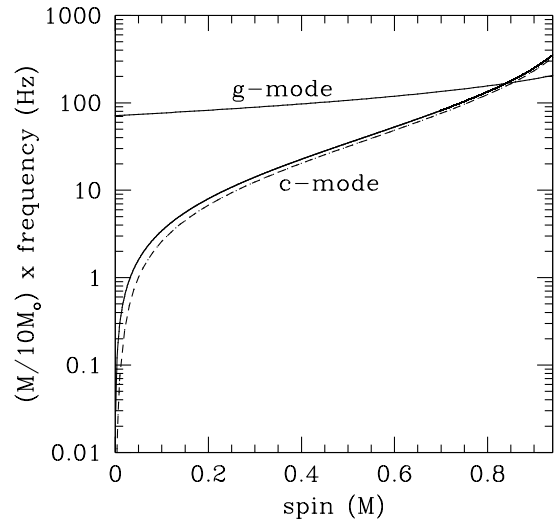,height=0.45\textwidth}
\psfig{figure=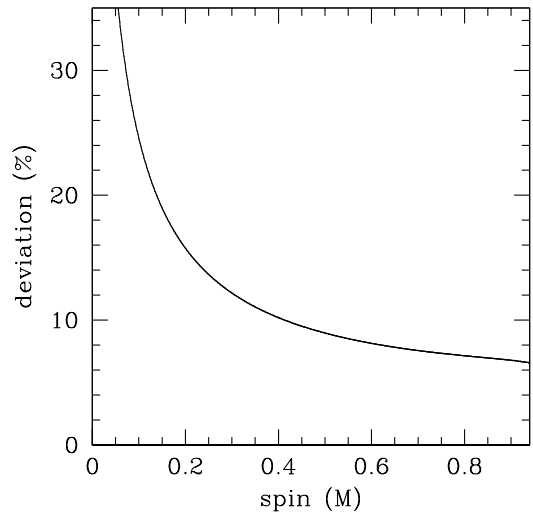,height=0.45\textwidth}
\end{center}
\caption{Top: Fundamental frequency of the $g$-mode and $c$-modes for a Kerr black hole as a function of spin. We evaluate the fundamental $c$-mode at the radius of the ISCO (solid line) and at the radius $r_c$ (dashed line). Bottom: Percent deviation between both expressions of the fundamental $c$-mode as a function of spin.}
\label{freqkerr}
\end{figure}

In Figure~\ref{freqkerr}, we plot as a function of the black hole spin (top panel) the fundamental frequencies of the $g$- and $c$-modes of a Kerr black hole using the above working definition (solid lines) as well as the fundamental $c$-mode (dashed line) evaluated at the radius $r_{\rm c}$ given by equation (\ref{rc}). The curves for the $c$-mode deviate primarily at lower spins. In Figure~\ref{freqkerr} we  also plot (bottom panel) this deviation between the expressions for the fundamental $c$-mode as a function of the spin $a$. In the range of spins that is relevant for this paper, i.e., $0\leq a/M \leq0.4$, the deviation is $\sim10-30\%$ except for very low values of the spin $a\lesssim0.1M$.

\begin{figure}[ht]
\begin{center}
\psfig{figure=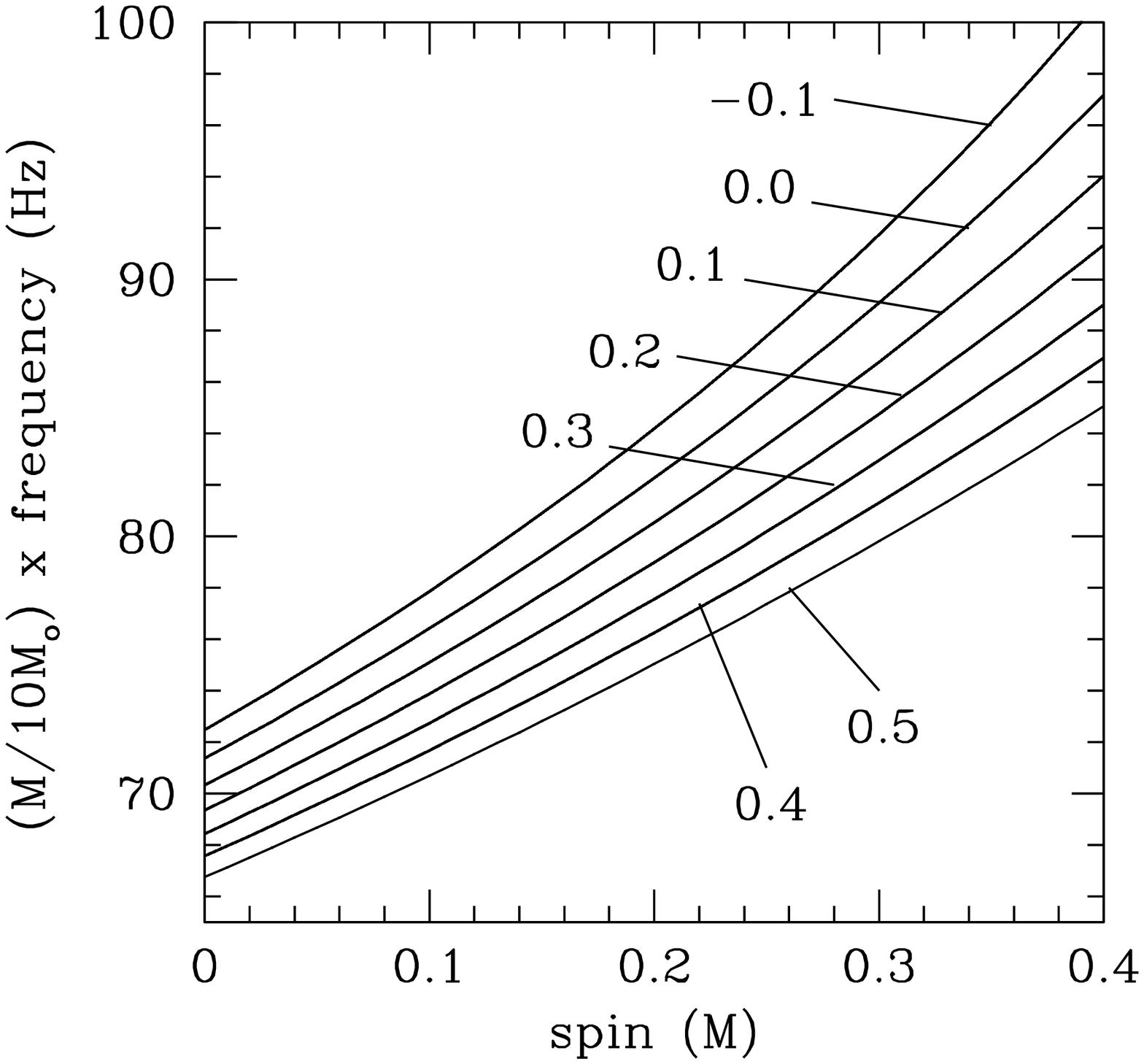,height=0.45\textwidth}
\psfig{figure=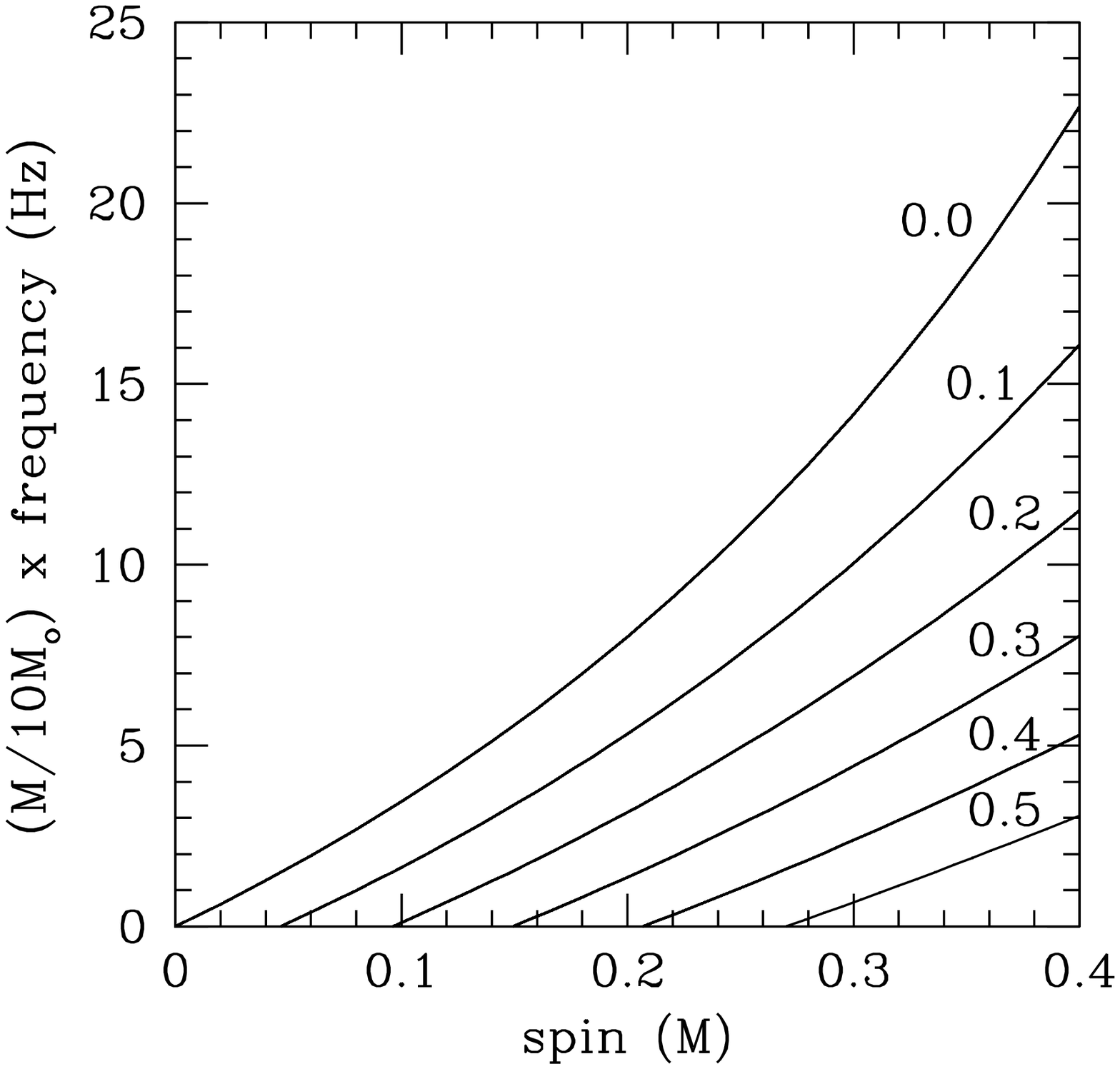,height=0.45\textwidth}
\end{center}
\caption{Frequencies of (top) the fundamental $g$-mode and (bottom) the fundamental $c$-mode scaled appropriately for a $10M_\odot$-mass black hole as a function of spin for several values of the parameter $\epsilon$. Both the fundamental $g$-mode and $c$-mode frequencies increase with increasing values of the spin but decrease with increasing values of the parameter $\epsilon$.}
\label{gcmodes}
\end{figure}

In Figure~\ref{gcmodes}, we plot (top panel) the fundamental $g$-mode and (bottom panel) the fundamental $c$-mode as a function of spin for several values of the parameter $\epsilon$ in units of a $10M_\odot$-black hole. In both cases, the frequency increases with increasing values of the spin, but decreases with increasing values of the parameter $\epsilon$.

\begin{figure}[ht]
\begin{center}
\includegraphics[width=0.45\textwidth]{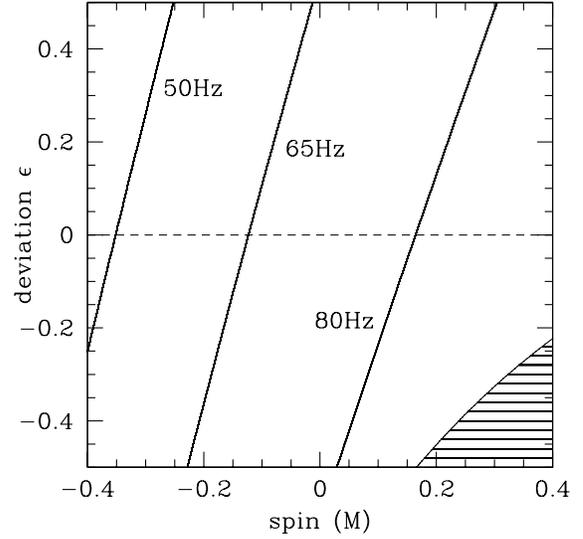}
\end{center}
\caption{Frequency contours of the fundamental $g$-mode as a function of the spin $a$ and the parameter $\epsilon$ scaled for a $10M_\odot$-black hole. The fundamental $g$-mode frequency increases with increasing values of the spin and decreasing values of the parameter $\epsilon$. The shaded region is excluded from the parameter space.}
\label{kappamax}
\end{figure}

In Figure~\ref{kappamax}, we plot frequency contours of the lowest order $g$-mode as a function of spin and the parameter $\epsilon$. The frequency of the fundamental $g$-mode increases for increasing values of the spin and decreasing values of the parameter $\epsilon$. The shaded region marks the excluded part of the parameter space.

\begin{figure}[ht]
\begin{center}
\includegraphics[width=0.45\textwidth]{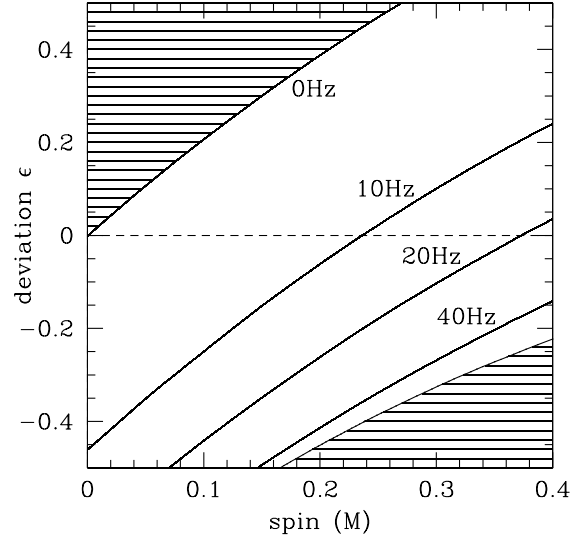}
\end{center}
\caption{Frequency contours of the fundamental $c$-mode as a function of the spin $a$ and the parameter $\epsilon$ scaled for a $10M_\odot$-black hole. The fundamental $c$-mode frequency increases with increasing values of the spin and decreasing values of the parameter $\epsilon$. The shaded region is excluded from the parameter space.}
\label{LTmax}
\end{figure}

As can be seen from Figure~\ref{kappamax}, contours of constant frequency of the fundamental $g$-mode are fairly linear in the spin $a$ and the parameter $\epsilon$. In order to further quantify this, we set $r\rightarrow r_0 + \delta a + \eta \epsilon$, where $r_0=8M$ is the location of the maximum radial epicyclic frequency for a Schwarzschild black hole and $\delta$ and $\eta$ are constants. Then, we expand the derivative of the radial epicyclic frequency given by expression (\ref{kappasquared}) with respect to radius around its maximum $r=r_0$ for $a=\epsilon=0$ and solve the equation
\begin{equation}
\left.\frac{d\kappa_{\rm r}^2}{dr}\right|_{r_0,a=\epsilon=0} + \left.\frac{d}{da}\frac{d\kappa_{\rm r}^2}{dr}\right|_{r_0,a=\epsilon=0}a + \left.\frac{d}{d\epsilon}\frac{d\kappa_{\rm r}^2}{dr}\right|_{r_0,a=\epsilon=0}\epsilon = 0
\end{equation}
for the parameters $\delta$ and $\eta$. We obtain the solutions
\begin{equation}
\delta=-\frac{47}{8\sqrt{2}},
\end{equation}
\begin{equation}
\eta=-\frac{5}{72}\left[34128\ln\left(\frac{4}{3}\right)-9835\right]M.
\end{equation}
From these expressions, we derive a linear approximation of the fundamental $g$-mode frequency in terms of the spin $a$ and the parameter $\epsilon$ given by the equation
\[
\kappa_{\rm r,max}^2=\frac{1}{M^2}
\]
\[
\times \left[ \frac{1}{2048} + \frac{15}{16384\sqrt{2}} \frac{a}{M} - \frac{ 5\left[ 5400\ln\left(\frac{4}{3}\right) - 1553 \right] }{ 16384 }\epsilon \right]
\]
\begin{equation}
\approx \left(5.09 + 6.75a^* - 1.54\epsilon\right)\times10^{3}\left( \frac{M}{10M_\odot} \right)^{-2}~{\rm Hz^2},
\end{equation}
where $a^*\equiv a/M$ . This approximation coincides with the maximum of the full expression of the radial epicyclic frequency given by equation (\ref{kappasquared}) to within 10\% across the full range of the values of the spin and the parameter $\epsilon$ considered in Figure~\ref{kappamax}.

In Figure~\ref{LTmax}, we plot frequency contours of the fundamental $c$-modes as a function of spin and the parameter $\epsilon$. The $c$-mode frequency rises with increasing values of the spin and decreasing values of the parameter $\epsilon$. The shaded regions mark the excluded parts of the parameter space. The top left corner is excluded according to our requirement that quasi-Kerr $c$-modes only exist in corotating disks. Since the Lense-Thirring frequency given by equation (\ref{omegalth}) is the difference between the epicyclic frequencies in the $\phi$ and $\theta$ coordinates, its main dependence is on the parameter $\epsilon$.

\begin{figure}[ht]
\begin{center}
\includegraphics[width=0.45\textwidth]{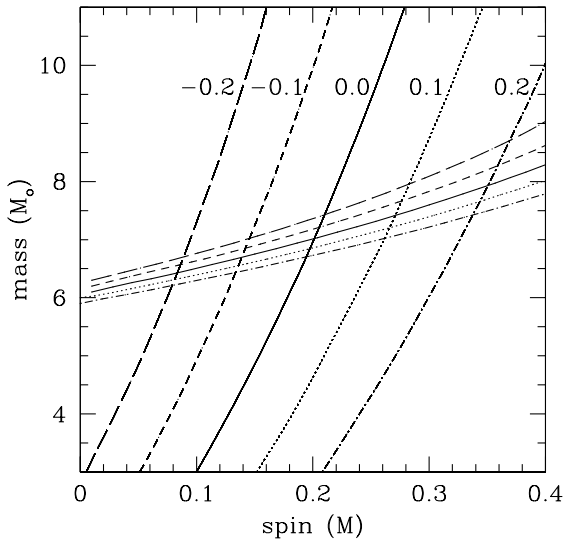}
\end{center}
\caption{Lines of constant $g$- and $c$-mode frequency (at 117.5Hz and 11.5Hz, respectively) as a function of mass and spin for several values of the parameter $\epsilon$. The intersection points of each corresponding pair of lines marks the particular combination of mass and spin for a given value of the parameter $\epsilon$.}
\label{m7a02}
\end{figure}

\begin{figure}[ht]
\begin{center}
\includegraphics[width=0.45\textwidth]{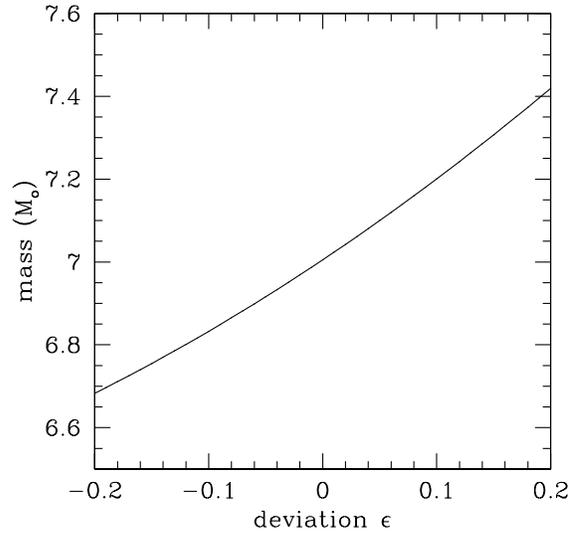}
\end{center}
\caption{Dependence of the black hole mass on the parameter $\epsilon$ determined by the intersection points as illustrated in Figure~\ref{m7a02}. Increasing values of the parameter $\epsilon$ correspond to a larger mass. A separate measurement of the mass breaks the degeneracy of the mass and the parameter $\epsilon$ allowing for a test of the no-hair theorem.}
\label{m7ep}
\end{figure}

In Figures~\ref{m7a02} and \ref{m7ep} we illustrate how measurements of $g$- and $c$-modes of a particular black hole can be used to test the no-hair theorem. For this purpose, we choose a black hole with a mass of $M=7M_\odot$ and a spin of $a=0.2M$. According to equations (\ref{kappasquared}) and (\ref{omegalth}), the corresponding frequencies of the $g$-mode and the $c$-mode, respectively, are 117.5Hz and 11.5Hz. In the following, we assume that these frequencies have been measured and identified as the respective lowest order $g$- and $c$-modes.

In Figure~\ref{m7a02}, we plot contours of constant $g$- and $c$-modes for frequencies of 117.5Hz and 11.5Hz, respectively, as a function of the black hole mass $M$ and spin $a$ for several values of the parameter $\epsilon$. The intersection point of each pair of lines marks the particular combination of the parameters $M$, $a$, and $\epsilon$ that is consistent with this hypothetical measurement. Increasing values of the parameter $\epsilon$ shift the intersection point to larger values of the mass and spin. In the case of a Kerr black hole, this reproduces the assumed values for the mass and spin.

Since we have three free parameters and two measurements, we can express the mass in terms of the parameter $\epsilon$. In Figure~\ref{m7ep}, we plot the mass as a function of the parameter $\epsilon$ for the intersection points given in Figure~\ref{m7a02}. Increasing values of the parameter $\epsilon$ correspond to a larger black hole mass. In combination with an independent mass measurement, the degeneracy between the mass and the parameter $\epsilon$ is broken, and all three parameters mass, spin, and quadrupole moment of the black hole can be determined. This, in turn, allows us to test the no-hair theorem via relation (\ref{kerrmult}).

\section*{3.2~~THE KINEMATIC RESONANCE MODEL}

\begin{figure}
\begin{center}
\includegraphics[width=0.45\textwidth]{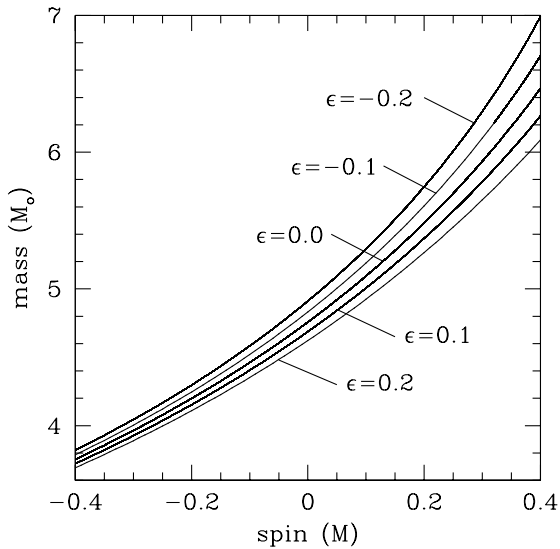}
\end{center}
\caption{Spin versus mass of a black hole for a QPO pair at frequencies 300~Hz and 450~Hz in 1:2 resonance for several values of the parameter $\epsilon$. The mass increases for increasing values of the spin and decreasing values of the parameter $\epsilon$.}
\label{res1}
\end{figure}

\begin{figure}
\begin{center}
\includegraphics[width=0.45\textwidth]{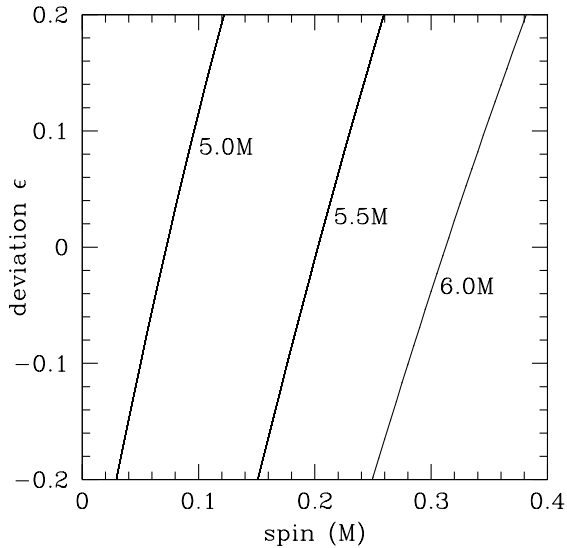}
\end{center}
\caption{Spin versus the parameter $\epsilon$ for the intersection points of the lines in Figure~\ref{res1} for different values of the mass $M$. The degeneracy between spin and the parameter $\epsilon$ can be broken by an independent spin measurement.}
\label{res2}
\end{figure}

QPOs can also be modeled as resonances among the Keplerian and epicyclic frequencies (Klu\'zniak \& Abramowicz 2001; Abramowicz et al. 2003). Several pairs of QPOs have been observed in galactic black holes at a ratio of $\approx3/2$ (see Remillard \& McClintock 2006). The assumption of either a 1:2 or a 1:3 resonance between the Keplerian frequency and the radial epicyclic frequency for the observed QPO pair in the binary GRO J1655-40 has lead to a determination of the spin of this black hole (Abramowicz \& Klu\'zniak 2001).

We demonstrate the potential of this approach to test the no-hair theorem by considering an example similar to GRO J1655-40. We follow the analysis in Abramowicz \& Klu\'zniak (2001), but we consider only the range of spins $-0.4\leq a/M \leq0.4$ in order to be consistent with our framework. For a 1:2 resonance between the Keplerian and radial epicyclic frequencies, we set $\Omega_{\rm \phi}=300~{\rm Hz}$ and $\kappa_{\rm r}=150~{\rm Hz}$. The resonance occurs at a radius $r_{\rm res}$ which is a function of mass and spin for a Kerr black hole and of mass, spin, and the parameter $\epsilon$ for a quasi-Kerr black hole.

In Figure~\ref{res1}, we plot the mass as a function of spin for  $\Omega_{\rm \phi}$ and $\kappa_{\rm r}$ in a 1:2 resonance for several values of the parameter $\epsilon$. The mass increases for increasing values of the spin and decreasing values of the parameter $\epsilon$. If the mass of the central object is known from an independent measurement, then only certain combinations of the spin and the parameter $\epsilon$ are consistent with such a measurement. For this illustration, we assume hypothetical mass measurements of $M=5.0,~5.5,~6.0M_\odot$, respectively.

In Figure~\ref{res2}, we plot the parameter $\epsilon$ as a function of spin for the intersection points of the lines in Figure~\ref{res1} with these hypothetical mass measurements. Higher values of the spin correspond to larger values of the parameter $\epsilon$ for a given black-hole mass. The degeneracy between spin and quadrupole moment can be broken by an independent spin measurement from, e.g., relativistically broadened iron lines (e.g., Brenneman \& Reynolds 2009), which in turn would uniquely determine the parameter $\epsilon$. In this QPO model, therefore, the no-hair theorem can be tested in conjunction with additional measurements of mass and spin.

\section{CONCLUSIONS}

In this series of papers, we are investigating a framework to test the no-hair theorem with observations of black holes in the electromagnetic spectrum (Paper~I) based on a quasi-Kerr metric (Glampedakis \& Babak 2006). Since, according to the no-hair theorem, mass and spin are the only free parameters of a black-hole spacetime, measurements of the mass, spin, and quadrupole moment can be used to test this fundamental property of black holes in general relativity.

In this paper, we derived expressions of the Keplerian and epicyclic frequencies in the quasi-Kerr spacetime and explored their properties. We showed that, for moderate values of the spin, the Keplerian frequency depends almost exclusively on the mass and spin of a given black hole, while the radial and vertical epicyclic frequencies depend significantly on the value of the parameter $\epsilon$ near the inner edge of the accretion disk in addition to their dependence on the mass and spin. We determined the location of the ISCO for the ranges of the parameters $-0.4\leq a/M \leq0.4$ and $-0.5 \leq \epsilon \leq 0.5$ and showed that the Lense-Thirring frequency for a particle at the ISCO changes by a factor of $\sim5$ for a black hole with spin $a=0.2M$ as the parameter $\epsilon$ is varied from $-0.2$ to $+0.2$. We discussed the emergence of two additional stable orbits inside the ISCO and possible consequences for astrophysical black holes.

We demonstrated how this approach can be applied to QPOs observed in galactic stellar-mass black holes and AGN focusing on two different models. If a pair of QPOs is observed and identified as the fundamental $g$- and $c$-modes, respectively, the mass, spin, and quadrupole moment of that black hole can be measured if the mass is known from dynamical observations. If the QPO pair is viewed as a nonlinear resonance between the Keplerian and epicyclic frequencies, the parameter $\epsilon$ can be measured together with independent measurements of the mass and spin.

In addition to galactic black holes, Sgr~A* is another prime target for tests of the no-hair theorem. High-resolution observations of stars in close orbit around Sgr~A* for over a decade have lead to a precise mass and distance measurement of Sgr~A* (Ghez et al. 2008; Gillessen et al. 2009). Recent VLBI observations resolved Sgr~A* on horizon scales (Doeleman et al. 2008) and pointed the way towards the first image of Sgr~A* in the near future (Fish \& Doeleman 2009). In Paper~II, we showed that a ring of light will surround the image of a black hole and that its shape depends directly on the mass, spin, quadrupole moment, and inclination of the black hole. In particular, we showed that the diameter depends predominantly on the mass, while the displacement and the asymmetry of this ring are proportional to the quantities $a\sin i$ and $\epsilon \sin^{3/2} i$, respectively, where $i$ is the inclination angle of the angular momentum of the black hole with respect to a distant observer. We noted that one additional observable is required to break the degeneracy of the displacement and the asymmetry with the inclination.

Here we argue that a full test of the no-hair theorem for Sgr~A* may be accomplished with a combination of VLBI imaging of the ring of light (or, more generally, the shadow; see Paper~II) and the observation of quasi-periodic variability in the emission from Sgr~A* using VLBI techniques (Doeleman et al. 2009). Such variability is thought to arise from density fluctuations orbiting around the center of mass. The orbital frequency of these inhomogeneities is the Keplerian frequency $\Omega_{\rm \phi}$. In this paper, we showed that, for moderate values of the spin, the dependence of $\Omega_{\rm \phi}$ on the parameter $\epsilon$ is negligible. Consequently, if high-frequency VLBI observations of Sgr~A* (Doeleman et al. 2009) can measure the Keplerian frequency of an orbiting hot spot and if Sgr~A* is not spinning rapidly, they will measure the spin of Sgr~A* irrespectively of the particular value of the parameter $\epsilon$. Therefore, measurements of the displacement and the asymmetry of the ring of light (or the shadow) of Sgr~A* will uniquely determine the inclination and the parameter $\epsilon$ allowing us to test the no-hair theorem.

We thank J. Schnittman and R. Wagoner for useful comments. This work was supported by the NSF CAREER award NSF 0746549.


\end{document}